\documentclass[12pt, a4paper]{article}

\usepackage[final]{showkeys}
\usepackage{a4}
\usepackage{amsmath}
\usepackage{amssymb}
\usepackage{epsfig}

\newcommand{\bel}[1]{\begin{equation}\label{#1}}
\newcommand{\bal}[1]{\begin{eqnarray}\label{#1}}

\newcommand{\expo}{\textrm{e}}
\newcommand{\imag}{\textrm{i}}

\begin{document}

\title{A variational approach to the partition function of an
       interacting many body system}

\author{Christian~Rummel and Helmut~Hofmann \\
\small\it{Physik-Department der Technischen Universit\"at M\"unchen,
          D-85747 Garching, Germany}}

\date{}

\maketitle

\begin{abstract}
For the calculation of the partition function $\mathcal{Z}$ of
small, isolated and interacting many body systems an improvement
with respect to previous formulations is presented. By including
inharmonic terms and employing a variational approach quantum
effects can be treated even at very low temperatures. In order to
test its accuracy the novel approach is applied to the exactly
solvable Lipkin-Meshkov-Glick model (LMGM). For thermodynamic
properties and level densities good agreement with the exact
calculations is found.
\end{abstract}

\section{Introduction}
\label{intro}

Level densities and thermodynamic properties of finite,
interacting many body systems are of greater interest in many
areas of physical research. A particular challenge has been the
temperature range where strong quantum effects render mean field
theories insufficient. Methods have been developed to approximate
the fundamental partition function of systems with a separable two
body interaction. They are especially valuable as an arbitrary
Hamiltonian with two body interaction $\hat{V}^{(2)}$ can always
be expanded into a series of separable terms of the form
\bel{twobodham-sum}
\hat{\mathcal{H}} =
\hat{T} + \hat{V}^{(2)} =
\hat{H} + \frac{1}{2} \sum_{\nu=1}^{n} k_{\nu}
\,\hat{F}^{\nu}\hat{F}^{\nu} \,,
\end{equation}
with one body operators $\hat{T}$, $\hat{H}$ and $\hat{F}^{\nu}$
(see e.g. \cite{ath.aly:npa:97}). Whereas the former must be
hermitian, the effective two-body interaction might also be of the form
$(\hat{F}^{\nu})^{\dagger} \hat{F}^{\nu}$. In nuclear physics the
quadrupole-quadrupole or the pairing interaction are examples  of
this type. As one knows pair correlations also play a role for
electronic systems, even if the latter are of mesoscopic scale as
e.g. for ultra-small super conducting metallic grains on which
substantial experimental \cite{nanopart} and theoretical
\cite{delft,rossignoli} research has focused recently.

A very elegant approach to the calculation of the partition function is
provided by the path integral formalism. Applying a Hubbard-Stratonovich
transformation (see e.g. \cite{negelej.orlandh}) the two-body
interaction may be turned into terms containing only one-body operators
and scalar auxiliary fields. In the simplest approximation the latter may be
treated at the mean field level, but extensions for the inclusion of many
body quantum fluctuations are possible. They can be accounted for
systematically in the sense of expansions about the semiclassical
approximation. A static approximation was originally developed
in order to study finite size effects in small superconductors
\cite{mub.scd.der:prb:72}. Later it had been applied in nuclear physics
calculating thermodynamic properties and level densities of hot nuclei
\cite{aly.zij:prc:84,lab.arp.beg:prl:88}. This method, which is valid in
the high temperature limit, was coined the "Static Path Approximation"
(SPA). Dynamical effects have then been included on the level of a local
RPA making use of Gaussian approximations on top of the static path.
Several version of this procedure exist under the names SPA+RPA
\cite{pug.bop.brr:ap:91}, PSPA (for "Perturbed Static Path
Approximation",  which we take over in the following)
\cite{ath.aly:npa:97}  and Correlated Static Path Approximation
(CSPA) \cite{ror.can:plb:97,ror.rip:npa:98}. These approaches are
applicable in a temperature range where quantum effects are no longer
negligible.

However, at a certain temperature $T_{0}$ the classical mean field
solution may become unstable in functional space and large
fluctuations render the Gaussian approximation insufficient. By
thoroughly analyzing the appearance of instabilities for the many
body problem at lower temperatures, it has been possible to go
beyond the PSPA achieving in this way a smooth behavior around
$T_{0}$. In this way the breakdown is shifted to the lower
temperature $T_{0}/2$ \cite{ruc.anj:epjb:02}. In the present paper
we want to report a novel approach \cite{rummel:phd} which
exploits a variational procedure. Its basic idea is taken over
from \cite{giachetti,fer.klh:pra:86} where the case of a particle
moving in a one dimensional potential has been treated. For the
many body systems to be studied here, the method consists of two
steps: First one goes beyond the Gaussian approximation by taking
into account fourth order inharmonic terms. Second the local RPA
modes are replaced by variational parameters. Their appropriate
values are found by minimizing the free energy. Formally this
method is applicable down to $T = 0$ and delivers very good
results for the free and internal energy of the many body system.
Unfortunately, at least for approximations adopted for our model
the specific heat exhibits some unphysical features at very low
temperatures; this problem will be addressed below in more detail.

The paper is organized as follows: After reviewing the conventional
approach to the partition function of interacting many body
systems in Sect.~\ref{partfunc} we develop the novel method in
detail in Sect.~\ref{sec-var}. In Sect.~\ref{LMG} it is tested at the
exactly solvable Lipkin-Meshkov-Glick model (LMGM)
\cite{lih.men.gla:np:65}. Finally the results are discussed and an
outlook of future work is given.

\section[Partition function]
{Partition function of an interacting many body system}
\label{partfunc}

In the sequel we concentrate on the schematic special case of just
one interaction term in (\ref{twobodham-sum}) and write
\bel{twobodham}
\hat{\cal H} = \hat{H} + \frac{k}{2} \,\hat{F}\hat{F} \,.
\end{equation}
Extension of the methods proposed in this paper to systems with
more (independent) collective degrees of freedom ($n > 1$) is
feasible in principle, although it may be tedious in detail. The
coupling constant $k = -|k|$ is taken to be negative. In the
nuclear case attractive interactions of this type generate
collective modes of iso-scalar nature \cite{bohra.mottelsonb.1}.
In the present paper we will restrict ourselves to such a situation.
With the path integral formulation repulsive interactions $(k > 0)$
leading to iso-vector modes have been studied in \cite{pug:prc:91}
within the SPA and in \cite{can.ror:prc:97} on the basis of SPA+RPA.
An extension of the methods proposed here to positive $k$ should be
possible extending the ideas of these references.
A detailed study will be undertaken in a future publication.

\subsection{General formalism}
\label{formalism}

The partition function of the grand canonical ensemble reads
\bel{Z}
\mathcal{Z}(\beta) =
\textrm{Tr} \,\exp \left( -\beta (\hat{\cal H} - \mu \hat{A}) \right) \,,
\end{equation}
where $\beta = 1/T$ is the inverse temperature (in units with
$k_{\textrm{B}} \equiv 1$), $\mu$ is the chemical potential and
$\hat{A}$ the number operator. This ${\cal Z}(\beta)$ shall be
evaluated by functional integrals in imaginary time
\cite{negelej.orlandh}. As mentioned before, the form of the
Hamiltonian in (\ref{twobodham}) invites one to introduce the mean
field approximation through a Hubbard-Stratonovich transformation
(HST). Thereby the separable two body interaction $\hat{F}\hat{F}$
is removed at the expense of introducing a collective variable
through an auxiliary path $q(\tau)$. Since this procedure is
well-known (see e.g. \cite{negelej.orlandh,ath.aly:npa:97}) we
simply state here the basic results which will serve as the
starting point for our analysis. (We will stick to the notation
used already in \cite{ruc.hoh:pre:01,ruc.anj:epjb:02} with the
exception of identifying $C \leftrightarrow \zeta$ .)

After introducing the Fourier expansion $q(\tau) = q_{0} + \sum_{r
\ne 0} q_{r} \,\exp (\imag\nu_{r}\tau)$ with amplitudes obeying
$q_{-r} = q_{r}^{*}$ and the Matsubara frequencies $\nu_{r} =
(2\pi/\hbar\beta) \,r = (2\pi T/\hbar) \,r$ the partition function
may be written in a form containing a static part that depends on
$q_0$ only and a factor $\zeta(\beta, q_{0})$ that corrects for
effects of the dynamics:
\bal{Z-athaly}
\mathcal{Z}(\beta) & = &
\sqrt{\frac{\beta}{2\pi |k|}} \int_{-\infty}^{+\infty}
\!\! dq_{0} \ \exp (-\beta \mathcal{F}^{\textrm{SPA}}(\beta, q_{0}))
\,\zeta(\beta, q_{0}) \nonumber \\
& = & \sqrt{\frac{\beta}{2\pi |k|}} \int_{-\infty}^{+\infty}
\!\! dq_{0} \ \exp (-\beta \mathcal{F}^{\textrm{eff}}(\beta, q_{0}))
\end{eqnarray}
Here $\mathcal{F}^{\textrm{eff}}(\beta, q_{0})$ is understood as
the effective free energy of the constituents of the total system
(\ref{twobodham}) moving in a time dependent mean field $q(\tau)$
of average $q_{0}$. The
\bel{FSPA}
\mathcal{F}^{\textrm{SPA}}(\beta,q_{0}) =
\frac{1}{2|k|} \,q_{0}^{2} - \frac{1}{\beta}
\,\textrm{ln} \,z(\beta,q_{0})
\end{equation}
is obtained in the limit where all time dependent fluctuations are
neglected. Both quantities, $\mathcal{F}^{\textrm{eff}}(\beta,
q_{0})$ and $\mathcal{F}^{\textrm{SPA}}(\beta, q_{0})$, should not
be mixed up with the free energy
\bel{freeen}
\mathcal{F}(\beta) = -\frac{1}{\beta} \ \textrm{ln} \,\mathcal{Z}(\beta)
\end{equation}
of the {\em total} self-bound system with Hamiltonian
(\ref{twobodham}). In (\ref{FSPA}) there appears the grand
canonical partition function $z(\beta, q_{0})$ which belongs
to the static part
\bel{1bHam-stat}
\hat{\mathcal{H}}_{\textrm{HST}}[q_{0}] = \hat{H} + q_{0} \,\hat{F}
\end{equation}
of the mean field approximation to the Hamiltonian
(\ref{twobodham}), which comes in via the HST. Obviously,
$\hat{\mathcal{H}}_{\textrm{HST}}[q_{0}]$ is a sum of only one
body operators; in the following its eigenenergies will be
denoted by $\epsilon_{l}(q_{0})$.

All contributions from the dynamical part of the auxiliary path
$q(\tau)$ are contained in the factor $\zeta(\beta,q_0)$ which can
formally be written as the functional integral
\bel{corrfactor} \zeta(\beta, q_{0}) = \int {\cal D}'q \ \exp
\left( -s_{\textrm{E}}(\beta, q_{0})/\hbar \right)
\end{equation}
(see e.g. \cite{ath.aly:npa:97} and \cite{ruc.anj:epjb:02}), where
a slightly different notation is used. Please note that the Fourier
representation of the measure ${\cal D}'q$ basically consists of
integrals over all $q_{r}$ but the $q_{0}$ itself.
To proceed further let us expand the Euclidean action
$s_{\textrm{E}}$ around the static path $q_{0}$:
\bel{defA}
s_{\textrm{E}} = \frac{\hbar\beta}{|k|} \left(
\sum_{r,s \ne 0} \lambda_{rs} \,q_{r} q_{s}
+ \sum_{r,s,t \ne 0} \rho_{rst} \,q_{r} q_{s} q_{t}
+ \sum_{r,s,t,u \ne 0} \sigma_{rstu} \,q_{r} q_{s} q_{t} q_{u} \right)
+ {\cal O}(q_{r}^{5})
\end{equation}
Here the coefficients are given by \cite{ruc.anj:epjb:02,rummel:phd}
\begin{eqnarray}
\lambda_{rs}(\beta, q_{0}) & = &
\frac{1}{2} \,\delta_{r,-s} + \frac{-|k|}{2! \,\beta}
\int_{0}^{\hbar\beta} d\tau_{r} d\tau_{s}
\ L_{rs}(\beta, q_{0}) \ , \label{deflambda} \\
\rho_{rst}(\beta, q_{0}) & = &
\frac{-|k|}{3! \,\beta}
\int_{0}^{\hbar\beta} d\tau_{r} d\tau_{s} d\tau_{t}
\ R_{rst}(\beta, q_{0}) \ , \label{defrho} \\
\sigma_{rstu}(\beta, q_{0}) & = &
\frac{-|k|}{4! \,\beta}
\int_{0}^{\hbar\beta} d\tau_{r} d\tau_{s} d\tau_{t} d\tau_{u}
\ [S_{rstu}(\beta, q_{0}) - 3 L_{rs}(\beta, q_{0})
L_{tu}(\beta, q_{0})] \label{defsigma}
\end{eqnarray}
with
\begin{eqnarray}
L_{rs}(\beta, q_{0}) & = &
\expo^{\imag\nu_{r}\tau_{r}} \expo^{\imag\nu_{s}\tau_{s}}
\ \frac{1}{\hbar^{2}} \,\langle \hat{\cal T} \hat{F}(\tau_{r})\hat{F}(\tau_{s})
\rangle_{q_{0}} \ , \label{defL} \\
R_{rst}(\beta, q_{0}) & = &
\expo^{\imag\nu_{r}\tau_{r}} \expo^{\imag\nu_{s}\tau_{s}}
\expo^{\imag\nu_{t}\tau_{t}}
\ \frac{1}{\hbar^{3}} \,\langle \hat{\cal T} \hat{F}(\tau_{r})\hat{F}(\tau_{s})
\hat{F}(\tau_{t}) \rangle_{q_{0}} \label{defR} \\
\textrm{and} \quad S_{rstu}(\beta, q_{0}) & = &
\expo^{\imag\nu_{r}\tau_{r}} \expo^{\imag\nu_{s}\tau_{s}}
\expo^{\imag\nu_{t}\tau_{t}} \expo^{\imag\nu_{u}\tau_{u}}
\ \frac{1}{\hbar^{4}} \,\langle \hat{\cal T} \hat{F}(\tau_{r})\hat{F}(\tau_{s})
\hat{F}(\tau_{t})\hat{F}(\tau_{u}) \rangle_{q_{0}} \ . \label{defS}
\end{eqnarray}
Similar expressions have to be evaluated for the coefficients of
higher order. In (\ref{defL}) to (\ref{defS}) $\hat{F}(\tau)$
denotes the operator $\hat{F}$ in the interaction picture.
$\hat{\cal T}$ is the ordering operator
in imaginary time and $\langle \ldots \rangle_{q_{0}}$ denotes the
expectation value with respect to the canonical density operator for the
Hamiltonian (\ref{1bHam-stat}). Notice that in this expansion no terms
survive which involve the $\tau$-independent factor $\langle
\hat{F}(\tau) \rangle_{q_{0}}$. Indeed, because of the presence of the
exponential factors the $\tau$-integrations make such terms vanish. For
this reason  in (\ref{defA}) terms linear in $q_r$ do not show up. The
expectation values for products of several $\hat{F}(\tau_r)$ can be
performed with the help of the Wick theorem. In this way these averages
are traced back to contractions. After carrying out the
$\tau$-integrations the expansion coefficients $\lambda_{rs}$,
$\rho_{rst}$ and $\sigma_{rstu}$ can be expressed by one body Green's
functions which are defined for the mean field Hamiltonian
(\ref{1bHam-stat}). Details of this procedure can be found in
\cite{ruc.anj:epjb:02,rummel:phd}.

\subsection[The conventional approach]
           {The conventional approach to quantum fluctuations}
\label{localRPA}

As mentioned earlier, within the SPA all dynamical contributions are
discarded, which formally is achieved by putting in (\ref{Z-athaly}) the
$\zeta^{\textrm{SPA}}(\beta, q_{0})$ equal to unity
\cite{mub.scd.der:prb:72,aly.zij:prc:84,lab.arp.beg:prl:88}.
For lower temperatures, when quantum properties tend to become
important, fluctuations can be incorporated within the conventional
version of the Perturbed Static Path Approximation (PSPA)
\cite{pug.bop.brr:ap:91}--\cite{ror.rip:npa:98},
\cite{ath.aly:npa:97,rossignoli,agb.ana:plb:98}.
There, the expansion (\ref{defA}) is truncated after the second order
terms in the $q_{r}$, which effectively means to describe quantum
effects on the level of local RPA. Using (\ref{deflambda}) and
(\ref{defL}) it can be shown that the following connection with the
response function $\delta\langle \hat{F} \rangle_{\omega} =
-\chi(\omega) \,\delta q(\omega)$ exists \cite{ruc.hoh:pre:01}:
\bel{lambdachi}
\lambda_{rs}(\beta, q_{0}) = \frac{1}{2} \,(1 + k
\chi(\imag\nu_{r})) \,\delta_{r,-s} = \frac{1}{2} \,\frac{1}{1 - k
\,\chi_{\textrm{coll}}(\imag\nu_{r})} \,\delta_{r,-s} = \frac{1}{2}
\,\lambda_{r}(\beta, q_{0}) \,\delta_{r,-s}
\end{equation}
In the second identity the general relation (see e.g.
\cite{bohra.mottelsonb.1}) $ \chi_{\textrm{coll}} = \chi(\omega)/(1 + k
\,\chi(\omega))$ of the intrinsic to the collective response function
has been used. Within the independent particle model (IPM) $\lambda_{r}$
can be written as
\bel{lambdachi-expl}
\lambda_{r}(\beta, q_{0}) =
\frac{\prod_{\mu} (\nu_{r}^{2} + \varpi_{\mu}^{2}(\beta, q_{0}))}
{\prod_{k>l}' (\nu_{r}^{2} + \omega_{kl}^{2}(q_{0}))} \,.
\end{equation}
Here, the $\varpi_{\mu}(\beta, q_{0})$ are the $M$ local RPA frequencies
which can be obtained from the secular equation
\cite{bohra.mottelsonb.1,hoh:pr:97}
\bel{secular}
1 + k\chi(\varpi_{\mu}) = 0 \,.
\end{equation}
The $\hbar\omega_{kl}(q_{0}) = \epsilon_{k}(q_{0}) -
\epsilon_{l}(q_{0})$ are the frequencies of the intrinsic excitations
and the primed product means that factors with $\epsilon_{k} =
\epsilon_{l}$ are omitted. Note that the numerator as well as the
denominator of the product (\ref{lambdachi-expl}) consists of $M$
factors. The largest contribution comes from the truly
collective modes, where the $\varpi_{\mu}$ differ essentially from the
corresponding $\omega_{kl}$. In PSPA all integrals to be evaluated in
(\ref{corrfactor}) are of Gaussian type, and cause no problem as long as
$\lambda_{r}(\beta, q_{0}) > 0$ for all $r \ne 0$. For systems with
instabilities one RPA frequency may become purely imaginary
$\varpi_{1}^{2} < 0$. Then the need for the
$\lambda_{r}(\beta, q_{0})$ to be positive implies a
condition for the temperature below which the conventional version of
the PSPA breaks down as soon as the first one, $\lambda_{1}(\beta,
q_{0})$, vanishes. This implies diverging $q_{1}$-integrals due to large
fluctuations in the $q_{1}$-direction. In the theories of dissipative
tunneling \cite{grabert,hap.tap.bom:rmp:90,weissu} (for many body
systems see \cite{ruc.hoh:pre:01,ruc.anj:epjb:02,rummel:phd})
the temperature $T_0$ at which this breakdown happens is known as the
``crossover temperature''. For all $T > T_0$ the PSPA is applicable
with the dynamical factor $\zeta^{\textrm{PSPA}}(\beta,q_{0})$ in
(\ref{Z-athaly}) being given by
\bel{defCPSPA}
\zeta^{\textrm{PSPA}}(\beta, q_{0}) =
\prod_{r>0} \frac{1}{\lambda_{r}(\beta, q_{0})} \,.
\end{equation}

In \cite{ruc.anj:epjb:02} the divergence of the $q_{1}$-integrals
at $T_{0}$ has been cured by taking into account in the
$s_{\textrm{E}}$ of (\ref{corrfactor}) inharmonic terms up to
fourth order in $q_{1}$ and $q_{2}$. The method developed there
and called ``extended PSPA'' (ePSPA) is very similar to the
treatment of the crossover region in dissipative tunneling
\cite{grabert}. More precisely, in the expansion (\ref{defA})
besides $\lambda_{rs}$ the three anharmonic coefficients
$\rho_{11-2}$, $\rho_{-1-12}$ and $\sigma_{11-1-1}$ have been
taken into account. The corresponding $q_{2}$- and
$q_{1}$-integrals can be evaluated analytically, resulting in a
lowering of the breakdown temperature down to $T_{0}/2$. At even
lower temperatures the fluctuations in directions besides the
$q_{1}$ become large, too. Due to the mutual coupling, in this
case a similar, analytical treatment is no longer possible. In a
further extension over the ePSPA, in \cite{ruc.anj:epjb:02} a Low
Temperature Approximation (LTA) has been proposed, that is
applicable at {\em any} temperature. To reach this result the
barrier region of $\mathcal{F}^{\textrm{SPA}}(\beta,q_{0})$ has
been excluded from the $q_{0}$-integral in (\ref{Z-athaly}). One
deficiency of the LTA is that most observables show an unphysical
sharp bend at $T = T_{0}$, which has its origin in this exclusion.

\section{A variational approach}
\label{sec-var}

Both the ePSPA and the LTA still have serious deficiencies in the low
temperature or the crossover region, respectively. Unfortunately, in the
same temperature range physically interesting effects  take place, like
e.g. some known phase transitions. For this reason a method is desirable
which delivers reliable results also there. In
\cite{giachetti,fer.klh:pra:86} a variational approach has been
developed for the simpler system of a particle moving in a one
dimensional potential, which allows one to evaluate the quantum
partition function at arbitrary temperature to high accuracy. Using the
formalism of coherent states, the underlying ideas have been applied to
many body systems in \cite{yos.kic.nak.noh:prc:00}. Here, we follow a
different approach and make direct use of the expansion (\ref{defA}). It
is useful to rewrite the functional integral (\ref{corrfactor}) for the
dynamical correction by introducing a {\em reference action}
$s_{\Omega}^{q_{0}}$:
\bal{corrfactor-var}
\zeta(\beta, q_{0})
& = & \int {\cal D}'q \ \exp [-s_{\Omega}^{q_{0}}/\hbar]
\ \exp [-(s_{\textrm{E}} - s_{\Omega}^{q_{0}})/\hbar] \nonumber \\
& = & \zeta_{\Omega}^{q_{0}}
\left\langle \exp [-(s_{\textrm{E}} - s_{\Omega}^{q_{0}})/\hbar]
\right\rangle_{\Omega}^{q_{0}}
\end{eqnarray}
Later this reference action will be chosen such that the functional
integral for the corresponding normalization factor
$\zeta_{\Omega}^{q_{0}} =
\int {\cal D}'q \ \exp \left( -s_{\Omega}^{q_{0}}/\hbar \right)$
can be evaluated exactly.
The average $\langle \ldots \rangle_{\Omega}^{q_{0}}$ used in
(\ref{corrfactor-var}) is taken with respect to $s_{\Omega}^{q_{0}}$.
Finally the inequality
\bel{JensenPeierls}
\left\langle \exp [-(s_{\textrm{E}} - s_{\Omega}^{q_{0}})/\hbar]
\right\rangle_{\Omega}^{q_{0}} \ge
\exp [-\langle s_{\textrm{E}} - s_{\Omega}^{q_{0}}
\rangle_{\Omega}^{q_{0}}/\hbar]
\end{equation}
will be employed, which derives from the convexity of the exponential
and can be taken over to functional integrals \cite{kleinerth}.
The average on the right hand side of (\ref{JensenPeierls}) is easier
to calculate than the one on the left hand side. Therefore the
strategy is to adjust a number of variational parameters in
$s_{\Omega}^{q_{0}}$ such that the right hand side is maximized.
Using the correction factor
\bel{zeta_var}
\zeta^\textrm{var}(\beta, q_{0}) =
\zeta_{\Omega}^{q_{0}}
\exp [-\langle s_{\textrm{E}} - s_{\Omega}^{q_{0}}
\rangle_{\Omega}^{q_{0}}/\hbar]
\end{equation}
this idea allows us to find an upper bound
$\mathcal{F}^{\textrm{var}}(\beta, q_{0}) \ge
\mathcal{F}^{\textrm{eff}}(\beta, q_{0})$ for the effective
free energy defined in (\ref{Z-athaly}) by {\em minimizing}
\bel{Feff}
\mathcal{F}^{\textrm{var}}(\beta, q_{0}) =
\mathcal{F}^{\textrm{SPA}}(\beta, q_{0})
- \frac{1}{\beta} \,\textrm{ln} \,\zeta^\textrm{var}(\beta, q_{0})
\end{equation}
with respect to the variational parameters.
The inequality (\ref{JensenPeierls}) implies the following
restrictions to the partition function (\ref{Z-athaly}) and the
free energy (\ref{freeen}) of the total system:
\bel{ineq-ZF}
\mathcal{Z}^\textrm{var}(\beta) \le \mathcal{Z}(\beta)
\qquad \textrm{and} \qquad
\mathcal{F}^\textrm{var}(\beta) \ge \mathcal{F}(\beta)
\end{equation}

A convenient form for $s_{\Omega}^{q_{0}}$ can be deduced from
the PSPA action by the replacement
\bel{replace}
s_{\textrm{E}}^\textrm{PSPA} =
\frac{\hbar\beta}{|k|}
\sum_{r>0} \lambda_{r}(\beta, q_{0}) \,|q_{r}|^{2}
\quad \longrightarrow \quad
s_{\Omega}^{q_{0}} =
\frac{\hbar\beta}{|k|}
\sum_{r>0} \Lambda_{r}(\beta, q_{0}; \Omega_{\mu}) \,|q_{r}|^{2} \,.
\end{equation}
Here for the stiffnesses $\Lambda_{r}$ in $q_{r}$-direction
several choices are convenient.
The two extreme cases are the following ones:
\bel{defLambda}
\Lambda_{r}(\beta, q_{0}; \Omega_{\mu}) =
\begin{cases} \begin{displaystyle}
\frac{\prod_{\mu}  (\nu_{r}^{2} + \Omega_{\mu}^{2})}
     {\prod_{k>l}' (\nu_{r}^{2} + \omega_{kl}^{2})}
\end{displaystyle}
& \textrm{(a)} \\[15pt]
\begin{displaystyle}
\frac{                 (\nu_{r}^{2} + \Omega_{1}^{2})
    \ \prod_{\mu\ne 1} (\nu_{r}^{2} + \varpi_{\mu}^{2})}
     {\prod_{k>l}'     (\nu_{r}^{2} + \omega_{kl}^{2})}
\end{displaystyle}
& \textrm{(b)}
\end{cases}
\end{equation}
In case (a) {\em all} $M$ local RPA frequencies $\varpi_{\mu}^{2}$
appearing in (\ref{lambdachi-expl}) are replaced by $m = M$
{\em adjustable parameters} $\Omega_{\mu}^{2}$.
In case (b) {\em only one} adjustable parameter $\Omega_{1}^{2}$ is
introduced ($m = 1$) which replaces that RPA mode $\varpi_{1}^{2}$,
which for systems with instabilities leads to the breakdown of the
PSPA below $T_{0}$. This greatly simplifies the variational procedure
by restricting the space spanned by the $\Omega_{\mu}^{2}$ to a one
dimensional one. It might also be possible to generalize to a
{\em small number} $1 < m \ll M$ of RPA frequencies,
namely those which belong to the modes with largest collectivity.

For the normalization factor $\zeta_{\Omega}^{q_{0}}$ defined in
(\ref{corrfactor-var}) the replacement
$\lambda_{r} \longrightarrow \Lambda_{r}$ leads to the following
expression (comp. (\ref{defCPSPA})):
\bel{C_Omega}
\zeta_{\Omega}^{q_{0}}(\beta, q_{0}; \Omega_{\mu}) =
\prod_{r>0} \frac{1}{\Lambda_{r}(\beta, q_{0}; \Omega_{\mu})}
\end{equation}
As we will see below, this form removes the convergence problems
of the PSPA even in regions where the $\mathcal{F}^{\textrm{SPA}}$
exhibits barriers.
The correction factor (\ref{zeta_var}) is evaluated best by
decomposing the action $s_{\textrm{E}}$ into
$s_{\textrm{E}} = s_{\textrm{E}}^{\textrm{PSPA}} +
\delta s_{\textrm{E}}$ and calculating
$\langle s_{\textrm{E}}^{\textrm{PSPA}} - s_{\Omega}^{q_{0}}
\rangle_{\Omega}^{q_{0}}$ and $\langle \delta s_{\textrm{E}}
\rangle_{\Omega}^{q_{0}}$ separately. With the definition
\bel{defPir}
\Pi_{r}(\beta, q_{0}; \Omega_{\mu}) =
\begin{cases} \begin{displaystyle}
\frac{\prod_{\mu} (\nu_{r}^{2} + \varpi_{\mu}^{2}) -
      \prod_{\mu} (\nu_{r}^{2} + \Omega_{\mu}^{2})}
     {\prod_{\mu} (\nu_{r}^{2} + \Omega_{\mu}^{2})}
\end{displaystyle}
& \textrm{(a)}
\\[15pt]
\begin{displaystyle}
\frac{\varpi_{1}^{2} - \Omega_{1}^{2}}
        {\nu_{r}^{2} + \Omega_{1}^{2}}
\end{displaystyle}
& \textrm{(b)}
\end{cases}
\end{equation}
the second order contribution is (see Appendix \ref{int2})
\bel{av2}
\langle s_{\textrm{E}}^{\textrm{PSPA}} - s_{\Omega}^{q_{0}}
\rangle_{\Omega}^{q_{0}} = \hbar \sum_{r>0} \Pi_{r} \,.
\end{equation}
The denominator of $\Pi_{r}$ is two orders higher in $\nu_{r} \sim r$
than the numerator. Therefore for large $r$ the terms fall off
sufficiently fast for the sum to converge.

In principle, when evaluating the
$\langle \delta s_{\textrm{E}} \rangle_{\Omega}^{q_{0}}$
one needs to know the expansion coefficients in (\ref{defA}) to
{\em all orders}. In the case of a particle in a one dimensional
potential this causes no problem as the coefficients can easily be
obtained from the potential $V(x)$ by mere differentiation
\cite{giachetti,fer.klh:pra:86}. For the many body
problem on the other hand,  more complicated expressions are needed,
like e.g. (\ref{defrho}) and (\ref{defsigma}) with
(\ref{defL})--(\ref{defS}). Due to symmetry arguments all contributions
to $\langle \delta s_{\textrm{E}} \rangle_{\Omega}^{q_{0}}$ vanish,
which have an odd number of $q_{r}$.  To get a first orientation of how
the novel variational approach works, we start from a truncated form
$s_{\textrm{E}}^{(4)} = s_{\textrm{E}}^{\textrm{PSPA}} +
\delta s_{\textrm{E}}^{(4)}$ of the action $s_{\textrm{E}}$ where
{\em only terms up to fourth order} are taken into account.
In Appendix \ref{int4} it is shown that the fourth order contribution
to the average reads
\bel{av4}
\langle \delta s_{\textrm{E}}^{(4)} \rangle_{\Omega}^{q_{0}} =
\frac{\hbar|k|}{\beta} \sum_{r,s>0} \sigma_{rs-r-s}
\,\frac{1}{\Lambda_{r}} \,\frac{1}{\Lambda_{s}} \,.
\end{equation}
The behavior of the coefficients $\sigma_{rs-r-s}$ for large
$r$ and $s$ is essential for the convergence of the sum (\ref{av4}).
Details of the calculation of $\sigma_{rstu}$ are given in
Sect.~4.1 of \cite{ruc.anj:epjb:02} and in \cite{rummel:phd}.
It turns out that $\sigma_{rs-r-s}$ consists of a number of
terms like
\bel{sigmabehav}
\sigma_{rs-r-s} \sim \frac{1}{\omega_{io} + \imag\nu_{r}}
                   \,\frac{1}{\omega_{ik} + \imag\nu_{-s}}
                   \,\frac{1}{\omega_{im} + \imag\nu_{r-s}} \,,
\end{equation}
which fall off sufficiently fast to insure the convergence of
(\ref{av4}).

Putting together all contributions up to fourth order
the correction factor (\ref{zeta_var}) can be written in the form
\bal{defCFKV}
\textrm{ln} \,\zeta^\textrm{(4)}(\beta, q_{0}) & = &
\textrm{ln} \,\zeta_{\Omega}^{q_{0}}
- \frac{1}{\hbar} \langle s_{\textrm{E}}^\textrm{PSPA} - s_{\Omega}^{q_{0}}
\rangle_{\Omega}^{q_{0}}
- \frac{1}{\hbar} \langle \delta s_{\textrm{E}}^{(4)} \rangle_{\Omega}^{q_{0}}
\\
& = & -\sum_{r>0} \textrm{ln} \Lambda_{r} - \sum_{r>0} \Pi_{r}
- \frac{|k|}{\beta} \sum_{r,s>0} \sigma_{rs-r-s}
\,\frac{1}{\Lambda_{r}} \,\frac{1}{\Lambda_{s}} \,.
\nonumber
\end{eqnarray}
This truncated form is used as an approximation to $\textrm{ln}
\,\zeta^\textrm{var}$ when minimizing $\mathcal{F}^{\textrm{var}}$
of (\ref{Feff}). The equations for the $\Omega_{\mu}^{2}$ do not
contain much physical information, for which reason they are not
shown explicitly. It may be noted, however, that without any
inharmonic terms ($\delta s_{\textrm{E}} = 0$) these equations are
solved by the RPA frequencies $\Omega_{\mu}^{2} =
\varpi_{\mu}^{2}$ in the temperature range $T > T_{0}$. For the
choice (b) in (\ref{defLambda}) and (\ref{defPir}) it is easy to
show that $\Omega_{1}^{2} = \varpi_{1}^{2}$ indeed determines the
only minimum of the corresponding $\mathcal{F}^{\textrm{var}}$. As
expected, under these restrictions the variational procedure
reproduces the PSPA.

\section[Application]{Application to the Lipkin-Meshkov-Glick model}
\label{LMG}

There are only few models for many body systems which allow for an exact
evaluation of quantities of interest. One of them is the
Lipkin-Meshkov-Glick model (LMGM) \cite{lih.men.gla:np:65}.
Using quasi-spin operators $\hat{J}_{x}$ and $\hat{J}_{z}$ and a
negative coupling constant $k = -|k|$ the Hamiltonian of the variant of
the LMGM used here reads
\bel{lip_basHam}
\hat{\cal H} = 2\epsilon \hat{J}_{z} + 2k \hat{J}_{x}^{2} \,.
\end{equation}
Identifying $\hat{H} = 2\epsilon \hat{J}_{z}$ and $\hat{F} =
2\hat{J}_{x}$ it has the structure of the Hamiltonian (\ref{twobodham})
with one separable two body interaction. The corresponding mean field
Hamiltonian (\ref{1bHam-stat}) reads $\hat{\cal H}_{\textrm{HST}}[q_{0}]
= 2\epsilon \hat{J}_{z} + 2q_{0} \hat{J}_{x}$. Its eigenvalues are
supposed to be $g$-fold degenerate. For the our calculations we took $g
= 10$ and $\epsilon = 5~\textrm{MeV}$.

\begin{figure}[htb] \begin{center}
\epsfig{file=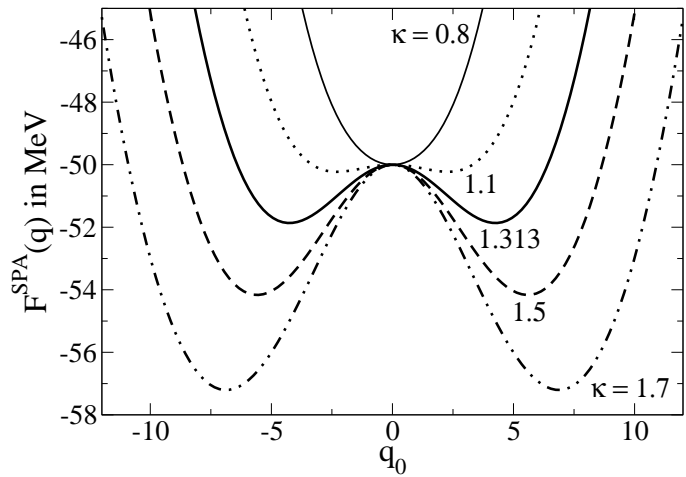, height=100mm, angle=-90}
\caption{\label{fig-FSPA}
$\mathcal{F}^{\textrm{SPA}}$ at $T = 0$ as a function of
$q_{0}$ for different effective coupling constants $\kappa$.}
\end{center} \end{figure}

\begin{figure}[p] \begin{center}
\epsfig{file=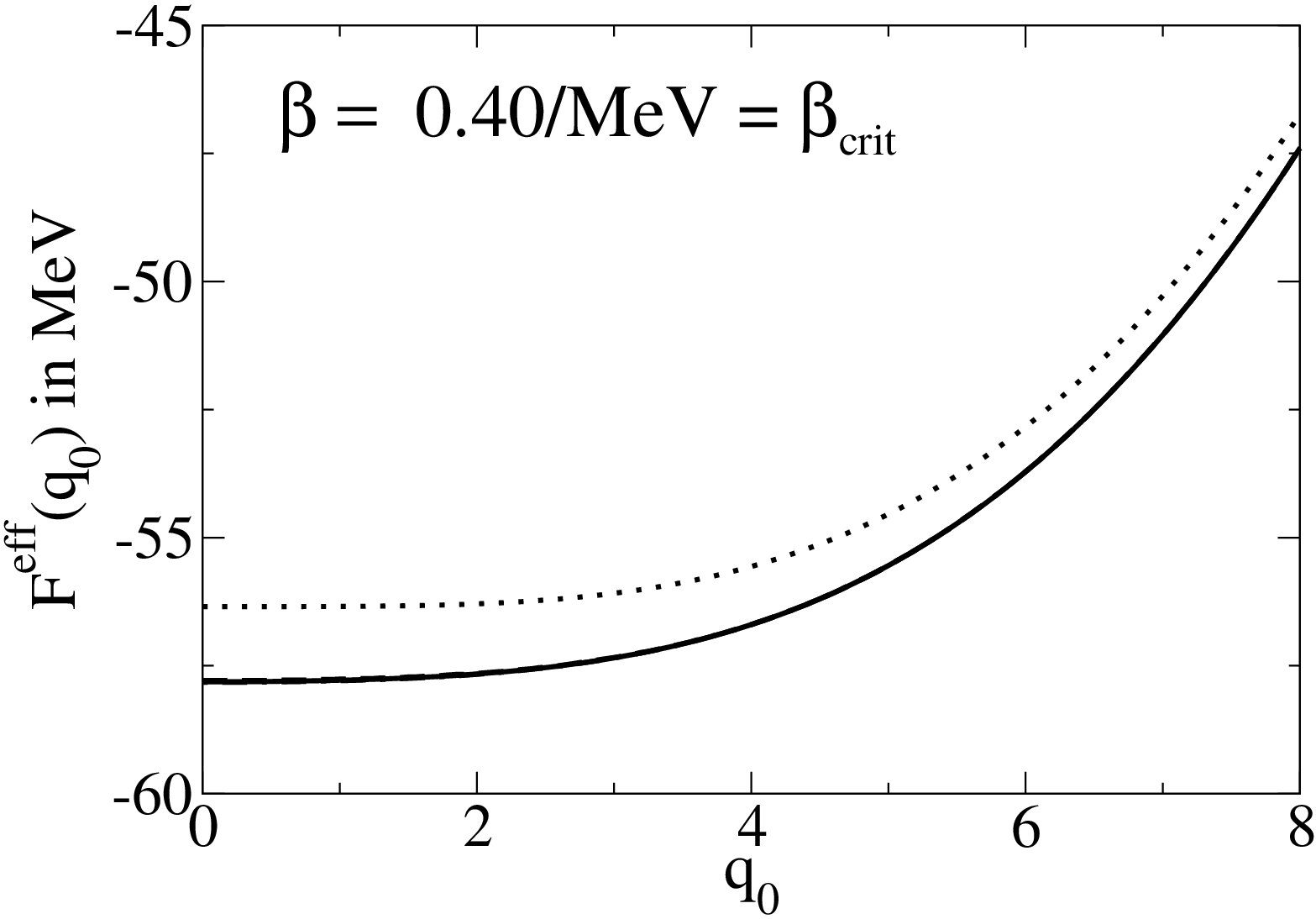, height=71mm, angle=-90}
\epsfig{file=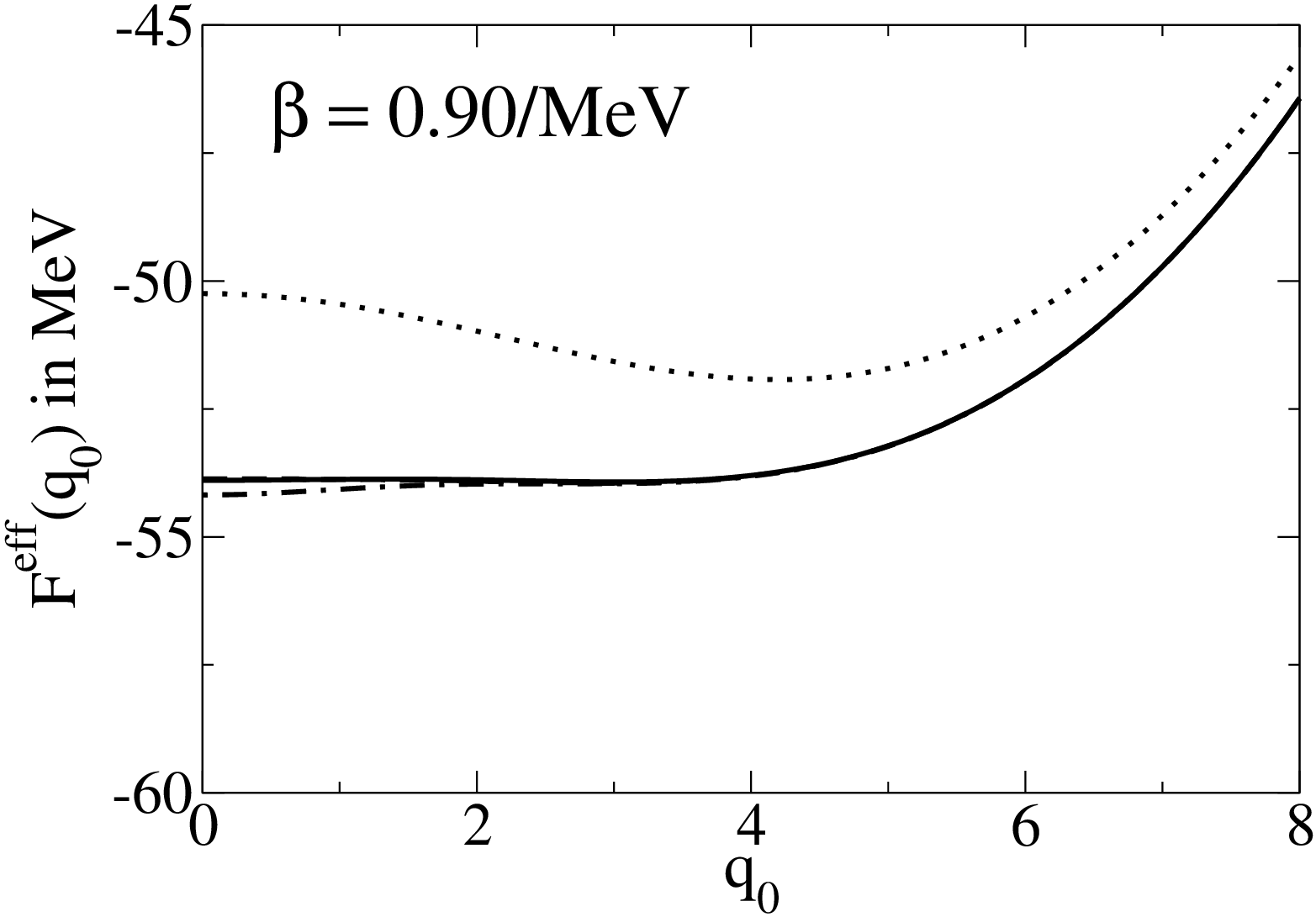, height=71mm, angle=-90}
\epsfig{file=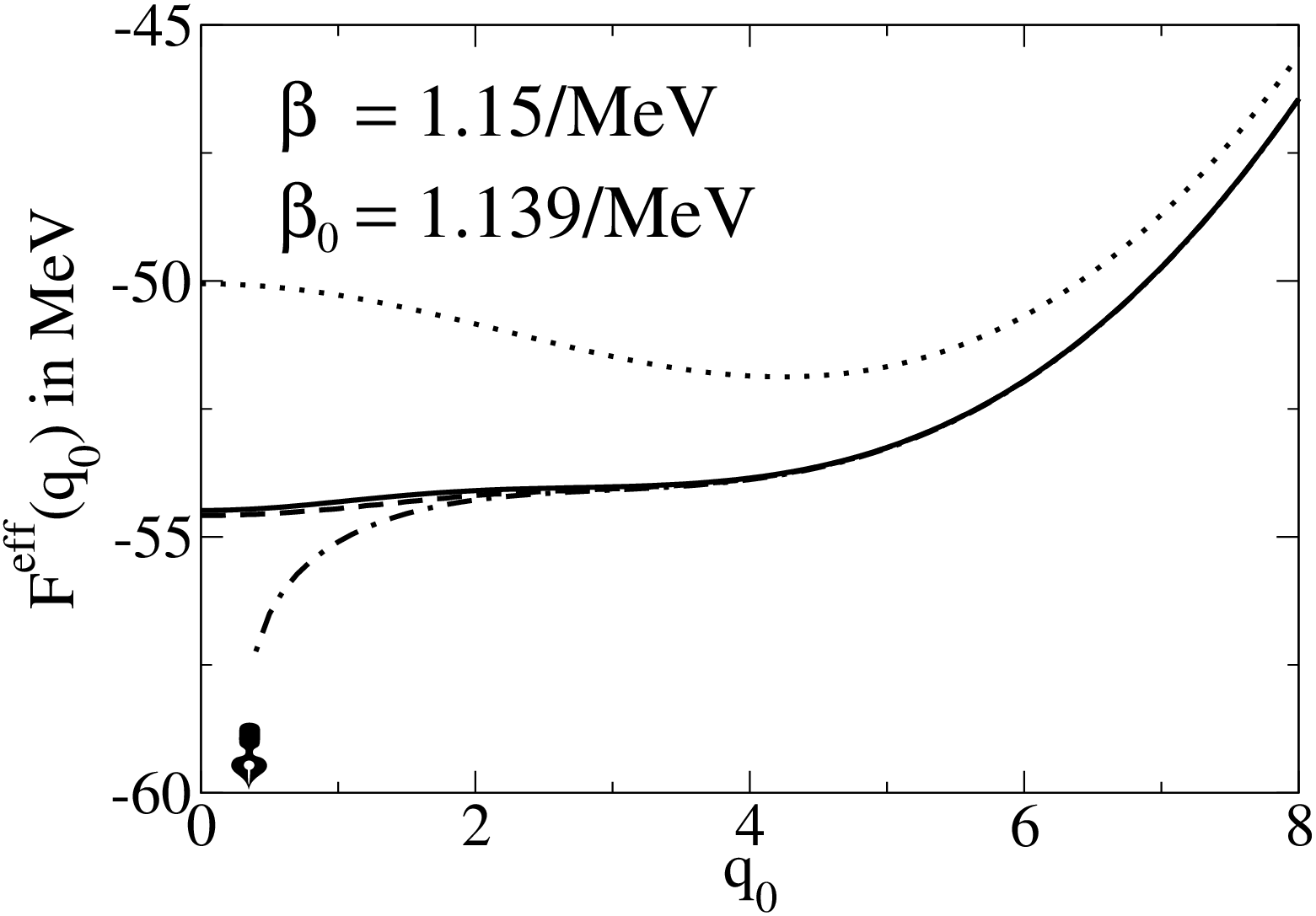, height=71mm, angle=-90}
\epsfig{file=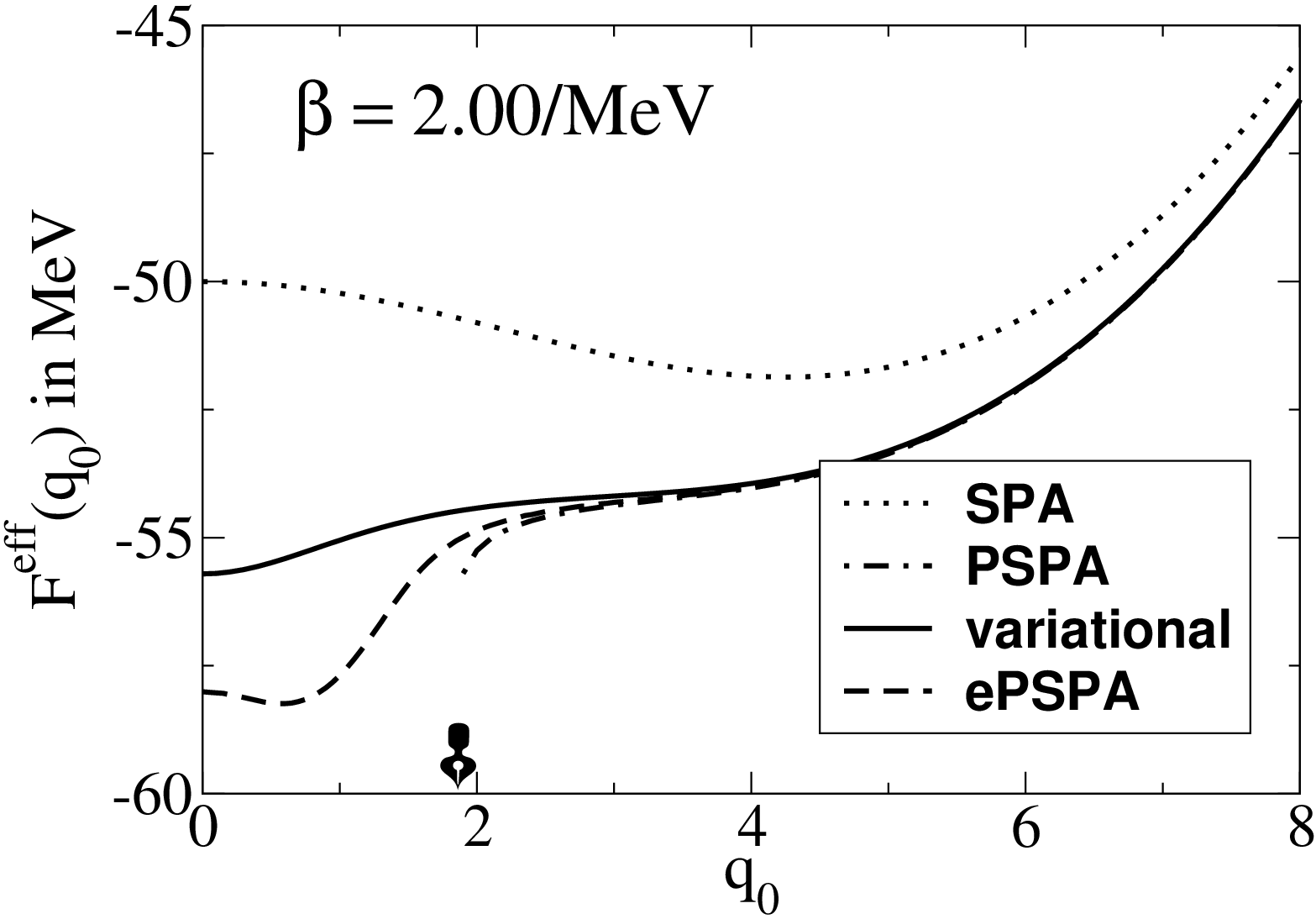, height=71mm, angle=-90}
\caption{\label{fig-FqLMG} $\mathcal{F}^{\textrm{SPA}}$ and different
approximations to the effective free energy $\mathcal{F}^{\textrm{eff}}$
of (\ref{Z-athaly}) for $\kappa = 1.313$ as a function of $q_{0}$ at
various temperatures. The inverse critical and crossover temperature are
$\beta_{\textrm{crit}} = 0.4~\textrm{MeV}^{-1}$ and $\beta_{0} =
1.139~\textrm{MeV}^{-1}$ respectively. $\mathcal{F}^{\textrm{eff}}$ is
an even function of $q_{0}$. The arrows point to the smallest $q_{0}$
where (at a given temperature) the PSPA still is formally applicable.}
\end{center} \end{figure}

This model has often been used to test the results of SPA type
approximations (see e.g.
\cite{pug.bop.brr:ap:91,ath.aly:npa:97,ror.rip:npa:98,ruc.anj:epjb:02}).
By varying the parameters of the model one gets a free energy
$\mathcal{F}^{\textrm{SPA}}(q_0) $ which shows the desired
property of developing a barrier at certain critical values, which may
or may not be washed out by temperature smoothing. This is best seen
after introducing the effective coupling constant
\bel{kappa}
\kappa = \frac{|k|g}{\epsilon} > 0 \,.
\end{equation}
As shown in Fig.~\ref{fig-FSPA}, at zero temperature the
$\mathcal{F}^{\textrm{SPA}}(q_0)$ of  (\ref{Z-athaly}) has a barrier at
$q_{0} = 0$  for all $\kappa > 1$ and none in the opposite case. The
influence of temperature on this functional form is demonstrated by the
dotted lines in Fig.~\ref{fig-FqLMG}, calculated for a $\kappa = 1.313$,
which was also used in \cite{pug.bop.brr:ap:91,ruc.anj:epjb:02}. The
barrier disappears whenever $\beta = 1/T$ becomes smaller than a
critical value $\beta_{\textrm{crit}} = 1/T_{\textrm{crit}}$. Another
property of the LMGM is that it has only one RPA mode $\varpi$.
Under such conditions the choices (a) and (b) of (\ref{defLambda})
and (\ref{defPir}) coincide and the optimization procedure of case
(a) is facilitated greatly.

Also shown in Fig.~\ref{fig-FqLMG} are the corrections to the
SPA which one gains by the approximations to the
$\mathcal{F}^{\textrm{eff}}$ of (\ref{Z-athaly}) described before.
For these calculations the number $N$ of Fourier coefficients
$q_{r}$ has been limited to $N = 100$, and hence also for the
products and sums in correction factors like (\ref{defCPSPA}) or
(\ref{defCFKV}). It has been checked that larger values of $N$ do
not change the results anymore. Considerable corrections to the
classical SPA are seen at any temperature for both the PSPA, the
ePSPA as well as the variational approach. Up to $\beta =
0.9~\textrm{MeV}^{-1} \approx 0.8 \beta_{0}$ differences among
them are hardly visible. The inverse crossover temperature for the
PSPA lies at $\beta_{0} = 1.139~\textrm{MeV}^{-1}$. At this
temperature the sign change of the coefficient $\lambda_{1}$
takes place at the barrier top at $q_{0} = 0$ of
$\mathcal{F}^{\textrm{SPA}}$ and leads to a divergence
of the $q_{1}$-integrals.
For $\beta > \beta_{0}$ this pathologic region grows to larger
$|q_{0}|$. Deviations between the ePSPA and the variational
approach are clearly visible at $\beta = 2~\textrm{MeV}^{-1}
\approx 1.8 \beta_{0}$.

It is remarkable that the effective free energy of the variational
approach (fully drawn line in Fig.~\ref{fig-FqLMG}) shows only little
structure in the barrier region of $\mathcal{F}^{\textrm{SPA}}$. This
feature does not change qualitatively at even larger $\beta$ and is in
agreement with the experience with the Feynman-Kleinert variational
approach (FKV) \cite{fer.klh:pra:86,kleinerth} if applied to a
one-dimensional double-well potential: There at low temperatures in the
approximation $W_{1}(x_{0})$ to the effective classical potential
$W(x_{0})$ (which can be seen as special one-dimensional cases of the
effective free energies $\mathcal{F}^{\textrm{var}}$ and
$\mathcal{F}^{\textrm{eff}}$, respectively) no barrier exists anymore.
In \cite{jaw.klh:cpl:87} Monte-Carlo simulations of the effective
classical potential $W(x_{0})$ have been discussed for the same problem.
It is reassuring that at small temperatures also such a $W(x_{0})$ was
found flat in the barrier region.

\begin{figure}[hbt] \begin{center}
\epsfig{file=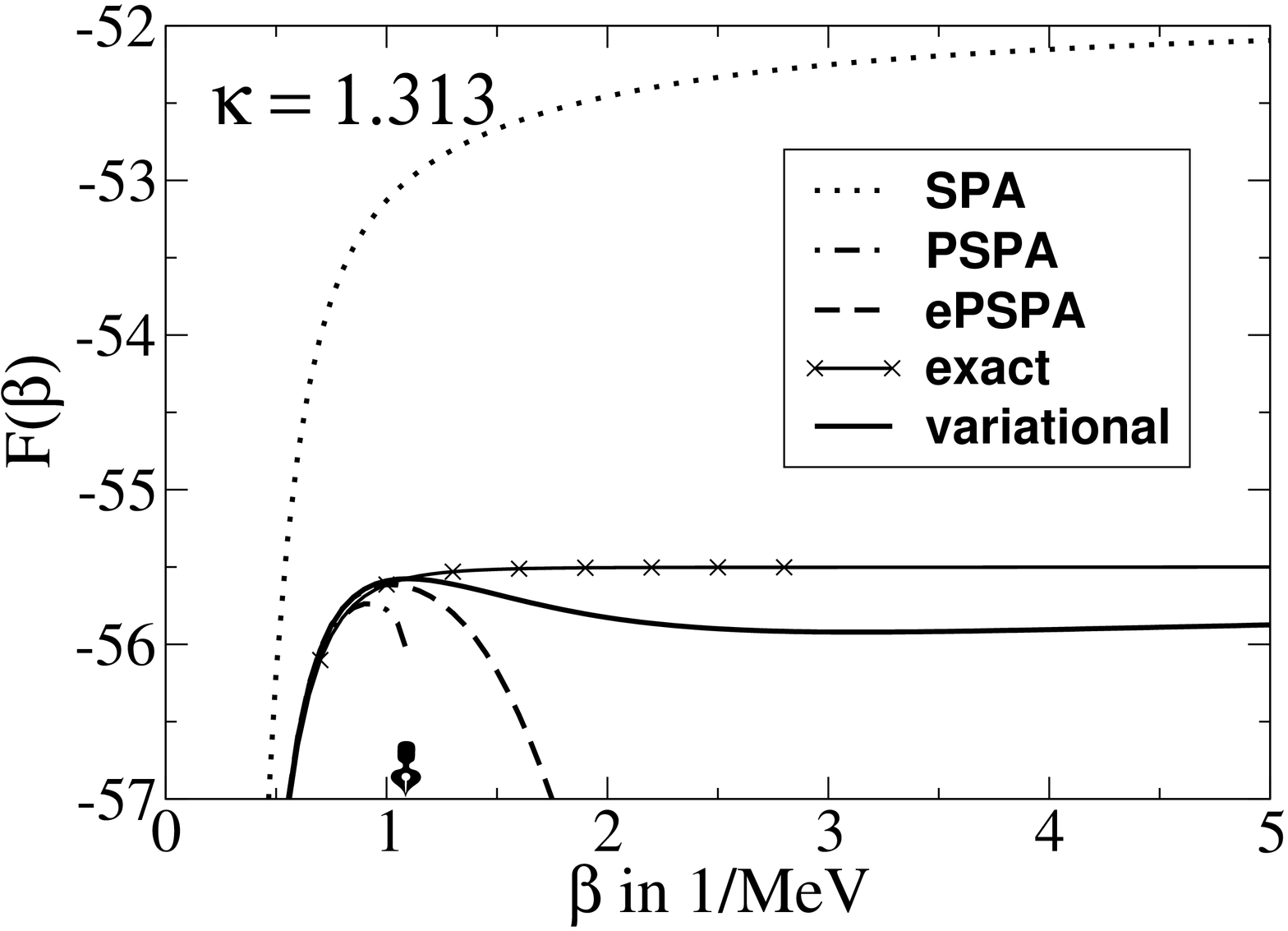, height=71mm, angle=-90}
\epsfig{file=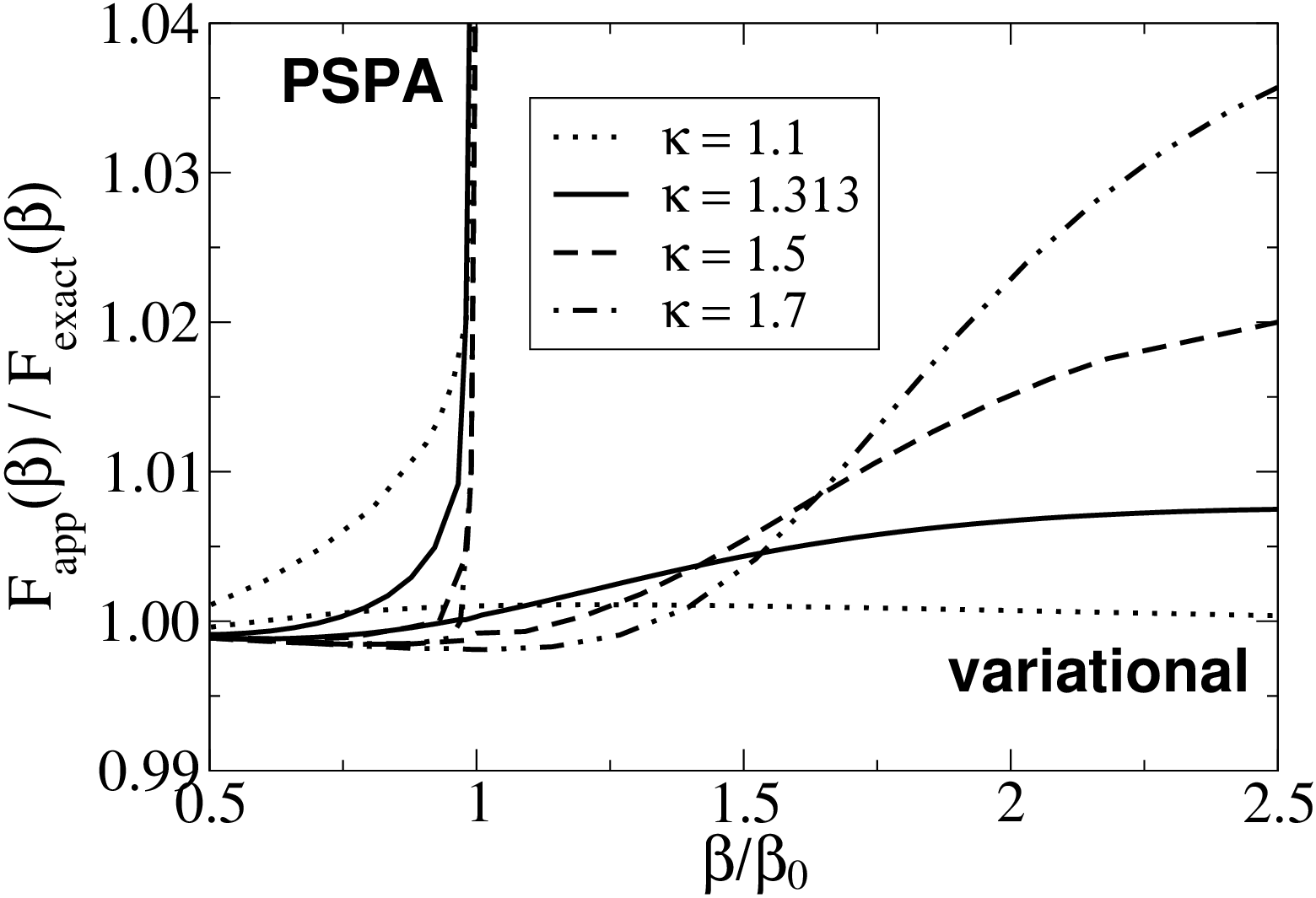,         height=71mm, angle=-90}
\caption{\label{fig-FbetaLMG-acc}
Left: Free energy of the total system in the LMGM as a function of
$\beta$. The arrow points to the inverse crossover temperature
$\beta_{0}$.
Right: Comparison of the accuracy of the PSPA and the variational
approach to the free energy of the LMGM.}
\end{center} \end{figure}
Next we turn to the free energy (\ref{freeen}) of the total system
with the partition function calculated from (\ref{Z-athaly}).
The quality of the different approximations are illustrated in
Fig.~\ref{fig-FbetaLMG-acc}. From the left panel it becomes
evident that for an effective coupling constant of  $\kappa =
1.313$ the classical SPA is acceptable only at high temperatures
(small $\beta$). In the regime $\beta < 0.9~\textrm{MeV}^{-1}
\approx 0.8 \beta_{0}$ the PSPA improves this classical result by
accounting for local RPA modes. A serious deficiency of the PSPA
is its breakdown at $\beta = \beta_{0} = 1.139~\textrm{MeV}^{-1}$.
Even in a region below $\beta_{0}$ its results are no longer
reliable. On the other hand the extension ePSPA of
\cite{ruc.anj:epjb:02} behaves completely regular in the crossover
region $\beta \approx \beta_{0}$. It can be used as a reasonable
approximation up to $\beta \approx 1.2~\textrm{MeV}^{-1} \approx
\beta_{0}$ but breaks down at still larger $\beta$. In contrast,
the variational approach is applicable even at very low
temperatures. There the relative error is of the order of 1\%
only.

Unfortunately, as can be seen from Fig.~\ref{fig-FbetaLMG-acc} the
free energy $\mathcal{F}^{(4)}(\beta)$ of the variational approach
is not a monotonously increasing function of $\beta$ as it should
be according to the first law of thermodynamics. However, this
deficiency is much worse for the PSPA and the ePSPA in a region $T
> T_{0}$ of formal applicability just before the breakdown.
To the best of our knowledge this problem has never been payed
attention to in the literature on PSPA. Conversely, the free
energy of our variational approach still is an increasing function
in the crossover region, where that of the PSPA and the ePSPA
already decrease. As we will explain below, there are indications
that the pathological feature of $\mathcal{F}^{(4)}(\beta)$ which
occurs at even lower temperatures is due to the truncation of the
reference action and not a property of the variational approach
itself. Also, it is seen that for $\beta > 1.2~\textrm{MeV}^{-1}
\approx \beta_{0}$ the variational free energy becomes smaller
than the exact result. Although this behavior seems to be in
contradiction to the restriction, in fact there is {\em no}
violation of (\ref{ineq-ZF}). To understand this, one should
notice that the inequality (\ref{JensenPeierls}) itself is valid
for the truncated action $s_{\textrm{E}}^{(4)}$ as well. Of
course, this inequality does not necessarily hold in case that
different actions are used on both sides, say like
$s_{\textrm{E}}$ on the left and $s_{\textrm{E}}^{(4)}$ on the
right. Hence, different to (\ref{ineq-ZF}) {\em no} inequality can
be written down between the approximation
$\mathcal{F}^{(4)}(\beta) \approx
\mathcal{F}^{\textrm{var}}(\beta)$ and the exact free energy
$\mathcal{F}(\beta)$.

We have been able to reproduce a similar behavior of the free
energy using the FKV of \cite{fer.klh:pra:86,kleinerth} to
calculate the partition function of a particle of mass $M$ moving
in a one-dimensional double-well potential $V(x) = -a x^{2} +
(1-a) x^{6}$ (where $0<a<1$) with anharmonicity of higher than
fourth order. Taking into account the inharmonic terms to all
orders the free energy of the FKV indeed is a monotonously
increasing function of $\beta$ and the inequality (\ref{ineq-ZF})
is fulfilled by the variational approach for all temperatures. A
truncation of the Euclidean action after the fourth order in the
$q_{r}$ leads to a free energy that is smaller than the exact
result and becomes decreasing at low temperatures. This
observation indicates that also the unphysical decrease of the
free energy $\mathcal{F}^{(4)}(\beta)$ in the left panel of
Fig.~\ref{fig-FbetaLMG-acc} does not reflect a deficiency of the
variational approach as such, rather it is an artefact of the
truncation of the reference action.

In the right panel of Fig.~\ref{fig-FbetaLMG-acc} the accuracy of
the PSPA and the variational approach is tested  at the example of
the free energy (\ref{freeen}) by showing the ratio of the
approximate results to the exact one. This is done for different
effective coupling constants $\kappa> 1$. It is observed that
within its range of applicability, which is to say for $\beta <
\beta_{0}$, the accuracy of the PSPA increases with increasing
$\kappa$. This may be traced back to the fact that the barrier
height of the effective free energy $\mathcal{F}^{\textrm{SPA}}$
at $T = 0$ grows with  $\kappa $, see again Fig.~\ref{fig-FSPA}.
However, the larger the barrier the smaller the probability for
quantum tunneling and, hence, the less important are fluctuations
of large scale. However, when these fluctuations become more and
more localized their treatment as local RPA modes becomes better
and better. Contrary to the PSPA, the accuracy of the variational
approach becomes worse the larger the barrier height. As can be
seen from the right panel of Fig.~\ref{fig-FbetaLMG-acc} this
feature is more pronounced for larger $\beta$. In the range $0 \le
\beta < \beta_{0}$, on the other hand, the  results of the
variational approach are better than those of the PSPA. One should
recall, perhaps, that nature of the inharmonic terms of
$\mathcal{F}^{\textrm{SPA}}$ taken into account by the variational
approach still is based on a local approximation. It does not base
on a large scale mode like the one necessary to describe quantum
tunneling correctly. These findings are in agreement with the
results of the FKV for the case of the one-dimensional double-well
potential: There, it has been observed that for high barriers the
FKV is unable to describe the splitting of low lying levels due to
tunneling correctly \cite{fer.klh:pra:86}.

\begin{figure}[htb] \begin{center}
\epsfig{file=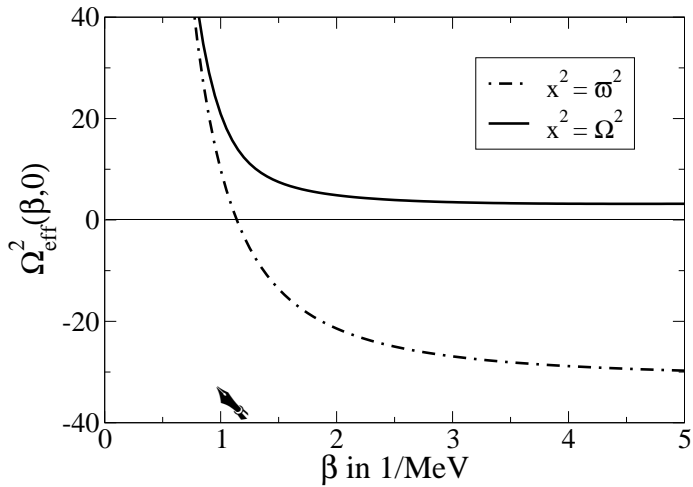, height=100mm, angle=-90}
\caption{\label{fig-freq}
Comparison of the ``stiffnesses'' $\Omega_{\textrm{eff}}^{2}$ of
$s_{\textrm{E}}^{\textrm{PSPA}}$ and $s_{\Omega}^{q_{0}}$
in $q_{1}$-direction defined in (\ref{critfreq}) using the
RPA-frequency $\varpi$ and the variational parameter $\Omega$,
respectively. The arrow points to the inverse crossover temperature.
The effective coupling strength is $\kappa = 1.313$.}
\end{center} \end{figure}
An interesting question is why the variational approach does not
suffer a real breakdown. For an answer it is useful to look at the
quantities
\bel{critfreq} \Omega_{\textrm{eff}}^{2}(\beta, q_{0}) = \left(
\frac{2\pi}{\hbar\beta} \right)^{2} + x^{2}(\beta, q_{0}) \qquad
\textrm{with} \qquad x^{2}(\beta, q_{0}) =
\begin{cases}
\varpi^{2}(\beta, q_{0}) \\[4pt]
\Omega^{2}(\beta, q_{0}) \,.
\end{cases}
\end{equation}
They largely define the stiffnesses of $s_{\textrm{E}}^{\textrm{PSPA}}$
and $s_{\Omega}^{q_{0}}$ in $q_{1}$-direction (see (\ref{replace}) and
mind $\nu_{1} = 2\pi/\hbar\beta$ as well as the fact that the product
over $\mu$ reduces to a single factor in the LMGM).
After the manipulations described in
Sect.~\ref{partfunc} and \ref{sec-var}, $\Omega_{\textrm{eff}}^{2}$
contributes to the denominators of the correction factors
(\ref{defCPSPA}) and (\ref{defCFKV}). Fig.~\ref{fig-freq} shows the
$\Omega_{\textrm{eff}}^{2}$ of  (\ref{critfreq}) as a function of
$\beta$ and for $q_{0} = 0$ (where the square of the RPA-frequency
$\varpi^{2}$ becomes negative for $\beta > \beta_{\textrm{crit}}$
and takes on the largest absolute value). For $x^{2} = \varpi^{2}$ the
$\Omega_{\textrm{eff}}^{2}$ becomes negative at the inverse
crossover temperature $\beta_{0}$ and the $q_{1}$-integral needed
in (\ref{corrfactor}) diverges at larger $\beta$. Consequently,
the formula (\ref{defCPSPA}), which has been derived under the
condition that {\em all} $q_{r}$-integrals converge, is no longer
valid under these circumstances. On the other hand using $x^{2} =
\Omega^{2}$ garanties $\Omega_{\textrm{eff}}^{2}$ to be {\em
strictly positive}. As now all $q_{r}$-integrals appearing in
(\ref{corrfactor-var}) are of Gaussian type, the correction factor
(\ref{defCFKV}) of the variational approach is well defined for
all $\beta$. As a result of this feature the effective free energy
$\mathcal{F}^{\textrm{var}}$ shows much less structure than
$\mathcal{F}^{\textrm{PSPA}}$, as already mentioned in connection
with Fig.~\ref{fig-FqLMG}.

Further tests of the accuracy of the different approximations may be
done for the internal energy $\mathcal{E}(\beta) = -\partial
\,\textrm{ln} \mathcal{Z}(\beta) /\partial \beta$ \cite{rummel:phd}  of
the total system or for the specific heat $\mathcal{C}(\beta) =
\beta^{2} \,\partial^{2} \,\textrm{ln} \mathcal{Z}(\beta) / \partial
\beta^{2}$. We concentrate on the latter for $\kappa > 1$ here.
\begin{figure}[htb] \begin{center}
\epsfig{file=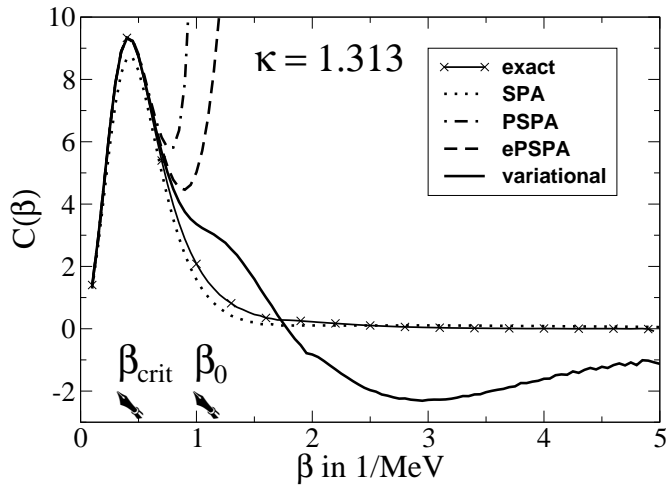, height=100mm, angle=-90}
\caption{\label{fig-ECbetaLMG}
The specific heat of the LMGM as a function of $\beta$ in various
approximations. The arrows point to the inverse critical temperature
$\beta_{\textrm{crit}}$ where the barrier in
$\mathcal{F}^{\textrm{SPA}}$ starts to build up at $q_{0} = 0$
and the inverse crossover temperature $\beta_{0}$.}
\end{center} \end{figure}
In Fig.~\ref{fig-ECbetaLMG} the classical SPA is seen to deliver a
{\em qualitatively correct} behavior of the specific heat at all
temperatures. For $\beta \approx \beta_{\textrm{crit}}$ it
under-estimates the exact result by about 10\%. At these smaller
$\beta$, however, the PSPA and the ePSPA supply much better
approximations. But they cease to be reliable already far below
their formal limits of applicability at $\beta_{0}$ and
$2\beta_{0}$ respectively. There the variational approach leads to
much better results, at least for not too large values of $\beta$.
For $\beta \gtrsim 1.7~\textrm{MeV} \approx 1.5\beta_{0}$, on the
other hand, it leads to negative values. Let us note that this
problem is much less pronounced for $\kappa < 1$ and closely
related to the appearance of positive curvatures of the
$\mathcal{F}(\beta)$ shown in the left panel of
Fig.~\ref{fig-FbetaLMG-acc}. The exact origin of this unphysical
behavior is still unclear and must be investigated further.
Perhaps and similar to the (apparent) violation of (\ref{ineq-ZF})
the effect might be related to the truncation of the expansion
(\ref{defA}) of the Euclidean action  $\langle\delta
s_{\textrm{E}} \rangle_{\Omega}^{q_{0}}$ at fourth order in
(\ref{av4}). However, studying analog truncations in
one-dimensional calculations within the FKV of
\cite{fer.klh:pra:86,kleinerth}, so far we have not found a
similar behavior of the specific heat even for systems with strong
inharmonic terms. Therefore the origin of the negative specific
heat in Fig.~\ref{fig-ECbetaLMG} will have to be investigated
further.

\begin{figure}[p] \begin{center}
\epsfig{file=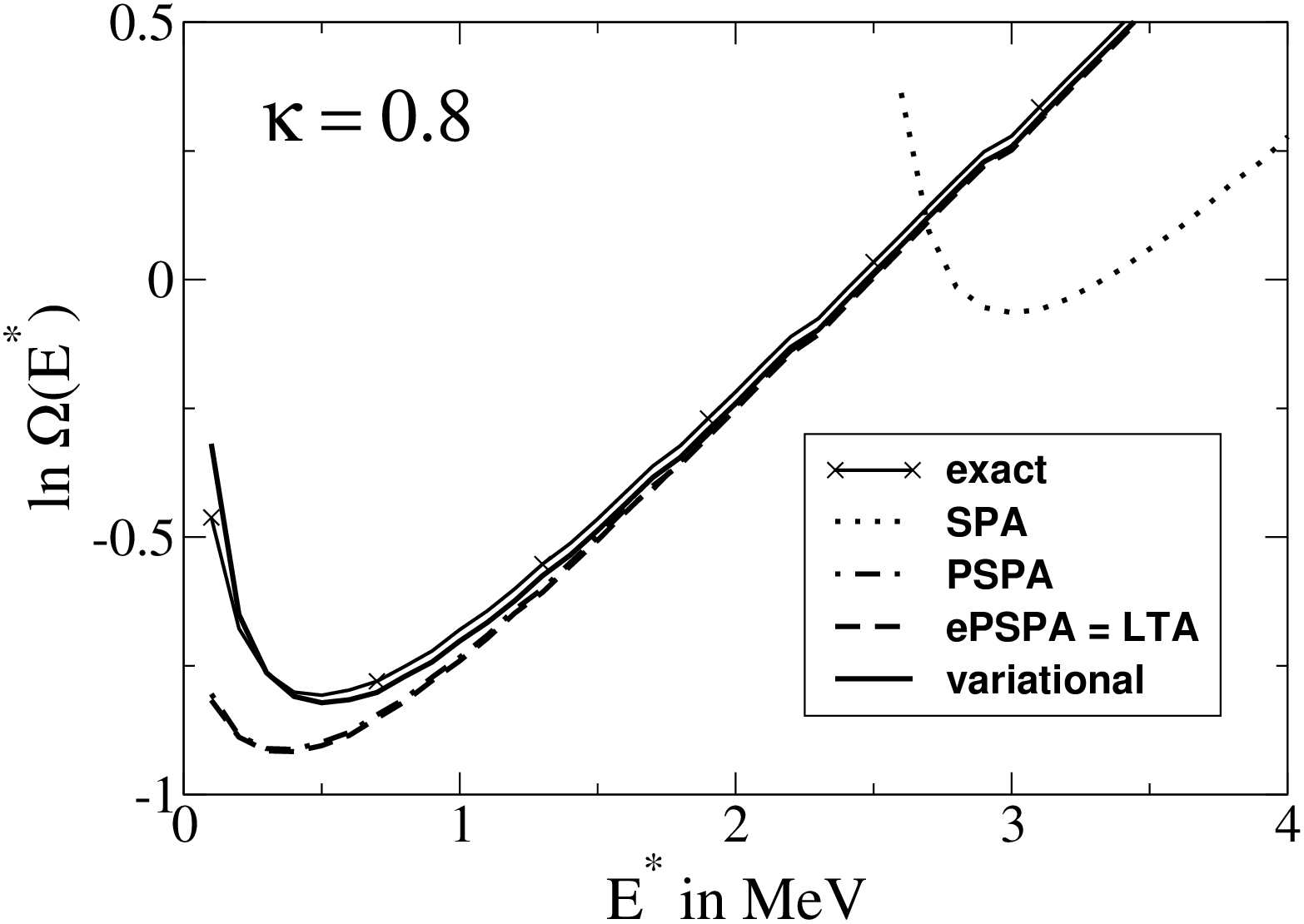,        height=100mm, angle=-90}
\epsfig{file=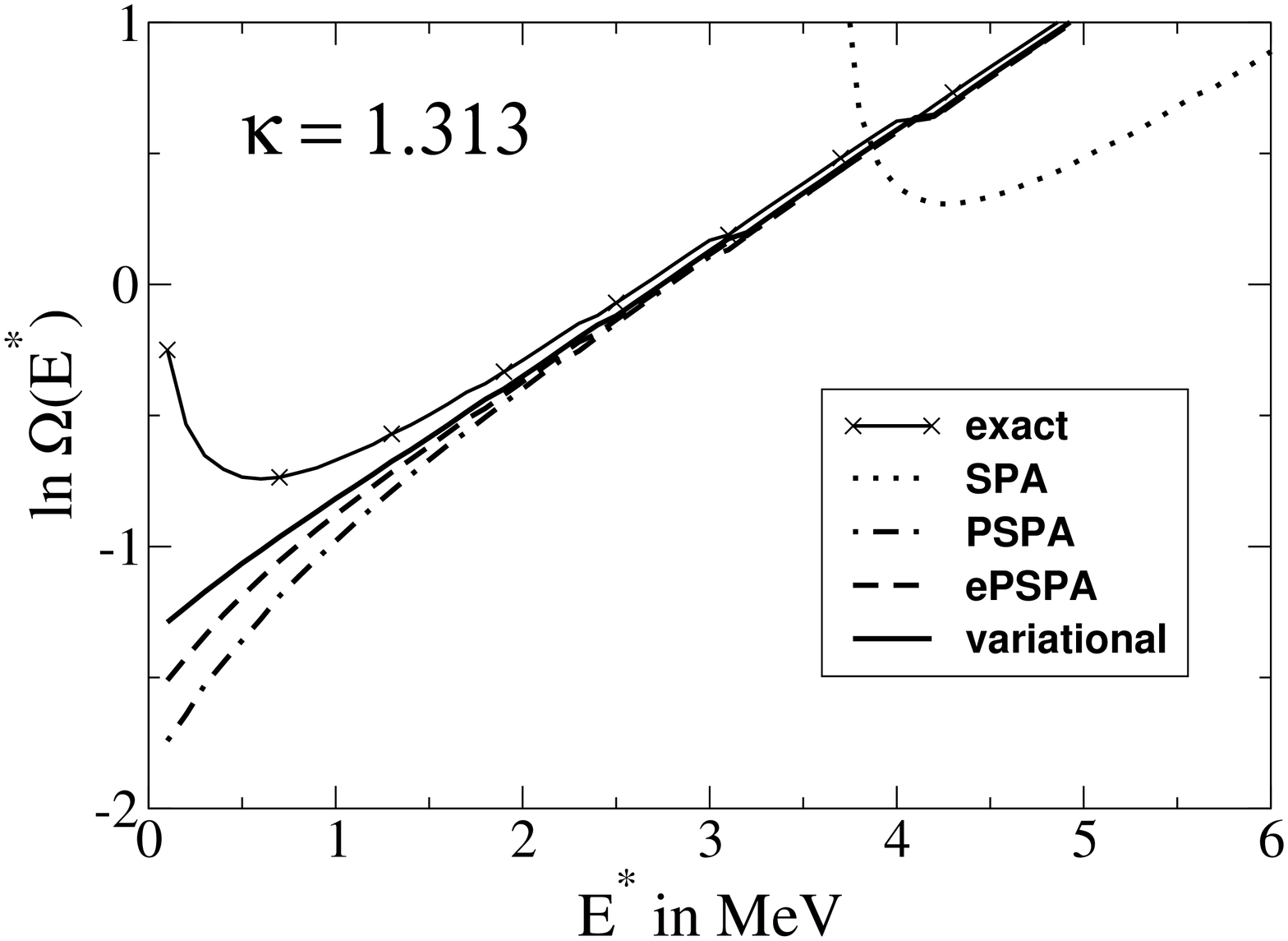, height=100mm, angle=-90}
\caption{\label{fig-rhoELMG} Logarithmic representation of the
Darwin-Fowler approximation to the level density as function of
the excitation energy $E^{*} = E - E_{0}$ in various
approximations. Whereas for $\kappa = 0.8$ no barrier is present
in $\mathcal{F}^{\textrm{SPA}}$, for $\kappa = 1.313$ it builds up
at $\beta = \beta_{\textrm{crit}}$.}
\end{center} \end{figure}
One prominent application of the SPA or the PSPA has been the
calculation of the level density $\rho$ of  finite, interacting
many body systems beyond the independent particle model
\cite{aly.zij:prc:84,lab.arp.beg:prl:88,pug.bop.brr:ap:91,agb.ana:plb:98}.
In Fig.~\ref{fig-rhoELMG} this level density is shown as function
of the excitation energy $E^{*} = E - E_{0}$. The calculations
have been performed with the Darwin-Fowler method (i.e. a saddle
point approximation to the inverse Laplace transform from
$\mathcal{Z}(\beta)$ to $\rho(E^{*})$, see e.g.
\cite{bohra.mottelsonb.1}). For the case of non-positive heat
capacities this becomes problematic as then the equation  $E =
\mathcal{E}(\beta)= -\partial \,\textrm{ln} \mathcal{Z}(\beta)
/\partial \beta$ no longer has unique solutions. In our
application we have concentrated on the one with smallest $\beta$,
simply because the other one lies in that regime of $\beta$ where
$\mathcal{E}(\beta)$ has a positive slope, implying
$\mathcal{C}(\beta) < 0$.  For excitations $E^{*}$ above about
$15~\textrm{MeV}$ (not shown in the plots) all approximations are
in good agreement with the Darwin-Fowler calculation which is
based on the exact partition function $\mathcal{Z}(\beta)$. At
smaller $E^{*}$, however, the level density of the classical SPA
is qualitatively wrong. Here, all other approximations lead to
considerable improvements over the SPA result. When the
$\mathcal{F}^{\textrm{SPA}}$ has no barrier, like for $\kappa =
0.8$ (see Fig.~\ref{fig-FSPA}), the PSPA and ePSPA can hardly be
distinguished from each other and show a qualitatively correct
behavior. On the other hand, for $\kappa = 1.313$, which is to say
when the $\mathcal{F}^{\textrm{SPA}}$ exhibits a barrier for low
temperature (see Fig.~\ref{fig-FqLMG}), both the PSPA as well the
ePSPA do not reproduce correctly the behavior of the calculation
using the full $\mathcal{Z}(\beta)$. Nevertheless, the ePSPA shows
a certain improvement over the PSPA. For $\kappa = 0.8$ the
variational approach reproduces perfectly well the calculation
done for the exact $\mathcal{Z}(\beta)$. For the case of a barrier
with $\kappa = 1.313$ one is in the situation discussed above
where within our present approximation the specific heat becomes
negative above a certain value of $\beta$, see
Fig.\ref{fig-ECbetaLMG}.  Although a clear improvement over the
ePSPA is seen, the present version of the  variational approach is
not capable of reproducing the qualitative behavior right at small
$E^{*}$. Finally we like to note that for $E^{*}$ below about $0.3
- 0.5$ MeV the Darwin-Fowler method itself becomes questionable.
This feature is in accord with the fact that the heat capacity
approaches zero at about $\beta \simeq 2 ~\text{MeV}^{-1}$.

\section{Discussion}
\label{disc}

A variational approach to the partition function
of a finite, interacting many body system has been developed and
applied to the exactly solvable LMGM. It has been shown that for
thermodynamic quantities like the free energy or the level density
it delivers quite accurate results.
This is especially important as the variational approach can formally
be applied without any problems in the low temperature regime
$\beta > \beta_{0}$, where the PSPA breaks down. In the crossover
range $\beta \approx \beta_{0}$ the variational approach considerably
improves the approximations used so far.

At very low temperatures there is still some deviation of the
variational from the exact results, which may be explained as
follows: Also the variational approach starts from a static auxiliary
field and takes into account {\em small} quantum fluctuations: $q(\tau)
\approx q_{0}$. At very low temperatures, however, {\em large scale}
fluctuations like quantum tunneling cannot be neglected. To include
such effects the full nonlinearity of the problem must be considered.
For interacting many body problems this is still an open problem.

The range of applicability of the method proposed is not limited to
problems of nuclear physics. Superconductivity in ultra-small metallic
grains \cite{nanopart} is a prime example for possible applications in
condensed matter physics, where up to now the path integral approach
has been limited to the PSPA \cite{rossignoli}.

Unfortunately, the benefits gained through the variational approach
must be payed for by a significantly larger computational effort.
The two quantities that most influence the computation time are the
number $N$ of Fourier coefficients $q_{r}$ (where $r = 1 \ldots N$)
and the dimension $m$ of the space spanned by the
variational parameters $\Omega_{\mu}^{2}$
(where $\mu = 1 \ldots m$ and $1 \le m \le M$).
With respect to the Fourier coefficients the computation time for
the PSPA and the ePSPA of \cite{ruc.anj:epjb:02} is limited by the
calculation of the $N$ coefficients $\lambda_{r}$ of
(\ref{lambdachi-expl}) (and the three coefficients
$\rho_{11-2}$, $\rho_{-1-12}$ of (\ref{defrho}) and
$\sigma_{11-1-1}$ of (\ref{defsigma}) in the case of ePSPA). In contrast,
for the variational approach in addition {\em all} the $N^{2}$
coefficients $\sigma_{rs-r-s}$ must be evaluated according to the
formulas given in \cite{ruc.anj:epjb:02,rummel:phd}. In the case of an
extension of the expansion (\ref{defA}) to order $2n$ even $N^{n}$
coefficients would have to be calculated. Please have in mind that the
one-dimensional analog \cite{giachetti,fer.klh:pra:86,kleinerth} of
the method proposed here does not suffer from this complication. There
the expansion coefficients can be read off directly from the Taylor
expansion of the underlying potential $V(x)$.

The computation time for the variational procedure required for
minimizing (\ref{Feff}) with (\ref{defCFKV}) is given by the task
of finding the absolute minimum of a scalar function in an
$m$-dimensional space.
As this effort strongly depends on the underlying function,
no strict statement is possible concerning the number of necessary
operations. Nevertheless it is intuitively clear that this number
strongly increases with the dimension $m$ of the corresponding space.
Therefore, the choice (a) in (\ref{defPir}) and (\ref{defLambda})
may imply serious drawbacks.
On the other hand, the choice (b) shows the way out of this
problem. The essential idea is to restrict the variational
procedure to the lowest-lying mode $\varpi_{1}^{2}$
that leads to the breakdown of the PSPA and contributes most to
the product of (\ref{lambdachi-expl}). A similar philosophy
is exploited in \cite{ruc.hoh:var-rate} in order to be able to
introduce dissipative effects. In cases where the restriction of
the number of the variational parameters to $m = 1$ is not good
enough, it should be possible to introduce variational parameters
$\Omega_{\mu}^{2}$ for a small number $m \ll M$ of such modes only
and use the RPA frequencies $\varpi_{\mu}^{2}$ for the rest.

\begin{appendix}

\section{Evaluation of the averages}
\label{averages}

\subsection{Second order contributions}
\label{int2}

Using the definitions (\ref{defLambda})--(\ref{defPir}) the second
order contribution to the average necessary on the right hand side of
(\ref{JensenPeierls}) is
\bal{avint2}
& & \langle s_{\textrm{E}}^\textrm{PSPA} - s_{\Omega}^{q_{0}}
\rangle_{\Omega}^{q_{0}} \nonumber \\
& = & \frac{1}{\zeta_{\Omega}^{q_{0}}} \int \prod_{s>0}
\frac{\beta}{\pi |k|} \,d\textrm{Re}(q_{s}) d\textrm{Im}(q_{s})
\,\exp \left( -\frac{\beta}{|k|} \sum_{s>0} \Lambda_{s} \cdot
\left( \textrm{Re}^{2}(q_{s}) + \textrm{Im}^{2}(q_{s}) \right) \right)
\times \nonumber \\
& & \frac{\hbar\beta}{|k|} \sum_{r>0} \Pi_{r} \Lambda_{r}
\,\left( \textrm{Re}^{2}(q_{r}) + \textrm{Im}^{2}(q_{r}) \right)
\nonumber \\
& = & \frac{\hbar\beta}{|k|} \sum_{r>0}
\Pi_{r} \Lambda_{r}^{2} \times \nonumber \\
& & \int \frac{\beta}{\pi |k|} \,d\textrm{Re}(q_{r}) d\textrm{Im}(q_{r})
\ \exp \left( -\frac{\beta}{|k|} \,\Lambda_{r} \cdot
\left( \textrm{Re}^{2}(q_{r}) + \textrm{Im}^{2}(q_{r}) \right) \right)
\,\left( \textrm{Re}^{2}(q_{r}) + \textrm{Im}^{2}(q_{r}) \right)
\nonumber \\
& = & \hbar \sum_{r>0} \Pi_{r} \,.
\end{eqnarray}
Here in the second identity we have carried out all Gaussian integrals
with $s \ne r$. In the third identity the remaining non-Gaussian
integrals are evaluated. Note that the sum in the final expression is
convergent due to the fact that the denominator of $\Pi_{r}$ is two
orders higher in $\nu_{r} = (2\pi/\hbar\beta) \,r$ than the numerator
(see (\ref{defPir})).

\subsection{Fourth order contributions}
\label{int4}

Due to symmetry arguments in fourth order the only contributing
terms read
\bal{avint4}
& & \langle \delta s_{\textrm{E}}^{(4)} \rangle_{\Omega}^{q_{0}}
\nonumber \\
& = & \frac{1}{\zeta_{\Omega}^{q_{0}}} \,\int \prod_{t>0}
\frac{\beta}{\pi |k|} \,d\textrm{Re}(q_{t}) d\textrm{Im}(q_{t})
\,\exp \left( -\frac{\beta}{|k|} \sum_{t>0} \Lambda_{t} \cdot
\left( \textrm{Re}^{2}(q_{t}) + \textrm{Im}^{2}(q_{t}) \right) \right)
\times \nonumber \\
& & \frac{\hbar\beta}{|k|} \sum_{r,s>0} \sigma_{rs-r-s}
\,(\textrm{Re}^{2}(q_{r}) + \textrm{Im}^{2}(q_{r}))
\,(\textrm{Re}^{2}(q_{s}) + \textrm{Im}^{2}(q_{s}))
\nonumber \\
& = & \frac{\hbar\beta}{|k|} \sum_{r,s>0} \sigma_{rs-r-s}
\,\Lambda_{r} \,\Lambda_{s} \times
\nonumber \\
& & \int \frac{\beta}{\pi |k|} \,d\textrm{Re}(q_{r}) d\textrm{Im}(q_{r})
\ \exp \left( -\frac{\beta}{|k|} \,\Lambda_{r} \cdot
\left( \textrm{Re}^{2}(q_{r}) + \textrm{Im}^{2}(q_{r}) \right) \right)
\left( \textrm{Re}^{2}(q_{r}) + \textrm{Im}^{2}(q_{r}) \right)
\nonumber \\
& & \int \frac{\beta}{\pi |k|} \,d\textrm{Re}(q_{s}) d\textrm{Im}(q_{s})
\ \exp \left( -\frac{\hbar\beta}{|k|} \,\Lambda_{s} \cdot
\left( \textrm{Re}^{2}(q_{s}) + \textrm{Im}^{2}(q_{s}) \right) \right)
\left( \textrm{Re}^{2}(q_{s}) + \textrm{Im}^{2}(q_{s}) \right)
\nonumber \\
& = & \frac{\hbar|k|}{\beta} \sum_{r,s>0} \sigma_{rs-r-s}
\,\frac{1}{\Lambda_{r}} \,\frac{1}{\Lambda_{s}} \,.
\end{eqnarray}
The quantities $\Lambda_{r}$ defined in (\ref{defLambda}) have been used.
The special case $r=s$ is included in (\ref{avint4}). Note that for
constant coefficients $\sigma_{rs-r-s}$ the sum would not
converge. Only the behavior (\ref{sigmabehav}) garanties convergence
of the sum.

\end{appendix}


\begin{thebibliography}{10}

\bibitem{ath.aly:npa:97}
H. Attias and Y. Alhassid, Nucl. Phys. {\bf A 625},  565  (1997).

\bibitem{nanopart}
D. Ralph, C. Black, and M. Tinkham, Phys.~Rev.~Lett. {\bf 74},  3241  (1995);
C. Black, D. Ralph, and M. Tinkham, Phys.~Rev.~Lett. {\bf 76},  688  (1996);
D. Ralph, C. Black, and M. Tinkham, Phys.~Rev.~Lett. {\bf 78},  4087  (1997).

\bibitem{delft}
F. Braun, J. v. Delft, D. Ralph and M. Tinkham, Phys.~Rev.~Lett.
{\bf 79}, 921 (1997);
F. Braun, J. v. Delft, Phys.~Rev.~Lett. {\bf 81}, 4712 (1998);
F. Braun, J. v. Delft, Phys.~Rev. {\bf B 59}, 9527 (1999).

\bibitem{rossignoli}
R. Rossignoli, J.P. Zagorodny, and N. Canosa, Phys.~Lett. {\bf A 258},  188
  (1999);
R. Rossignoli, N. Canosa, P. Ring, Ann.~Phys. {\bf 275}, 1 (1999);
N. Canosa and R. Rossignoli, Phys.~Rev. {\bf B 62},  5886  (2000);
R. Rossignoli and N. Canosa, Phys.~Rev. {\bf B 63},  134523  (2001).

\bibitem{negelej.orlandh}
J.~W. Negele and H. Orland, {\em Quantum Many-Particle Systems}
  (Addison-Wesley, Reading, MA, 1988).

\bibitem{mub.scd.der:prb:72}
B. M\"uhlschlegel, D. Scalapino, and R. Denton, Phys.~Rev. {\bf B 6},  1767
  (1972).

\bibitem{aly.zij:prc:84}
Y. Alhassid and J. Zingman, Phys.~Rev. {\bf C 30},  684  (1984).

\bibitem{lab.arp.beg:prl:88}
B. Lauritzen, P. Arve and G.F. Bertsch, Phys.~Rev.~Lett. {\bf 61},
2835 (1988).

\bibitem{pug.bop.brr:ap:91}
G. Puddu, P.~F. Bortignon, and R.~A. Broglia, Ann. Phys. (San Diego) {\bf 206},
   409  (1991).

\bibitem{ror.can:plb:97}
R. Rossignoli and N. Canosa, Phys. Lett. {\bf B 394},  242  (1997).

\bibitem{ror.rip:npa:98}
R. Rossignoli and P. Ring, Nucl.~Phys. {\bf A 633},  613  (1998).

\bibitem{ruc.anj:epjb:02}
C. Rummel and J. Ankerhold, Eur.~Phys.~J. {\bf B 29},  105  (2002).

\bibitem{rummel:phd}
C. Rummel, Ph.D. thesis, Technische Universit\"at M\"unchen, 2004.

\bibitem{giachetti}
R.~Giachetti and V.~Tognetti, Phys.~Rev.~Lett. {\bf 55}, 912 (1985);
R.~Giachetti and V.~Tognetti, Phys.~Rev. {\bf B 33}, 7647 (1986).

\bibitem{fer.klh:pra:86}
R.~P. Feynman and H. Kleinert, Phys. Rev. {\bf A 34},  5080  (1986).

\bibitem{lih.men.gla:np:65}
H.~J. Lipkin, N. Meshkov, and A.~J. Glick, Nucl. Phys. {\bf 62},  188  (1965).

\bibitem{bohra.mottelsonb.1}
A. Bohr and B.~R. Mottelson, {\em Nuclear Structure} (Benjamin, London, 1975),
  Vol.~1.

\bibitem{pug:prc:91}
G. Puddu, Phys.~Rev. {\bf C 44},  905  (1991).

\bibitem{can.ror:prc:97}
N. Canosa and R. Rossignoli, Phys.~Rev. {\bf C 56},  791  (1997).

\bibitem{ruc.hoh:pre:01}
C. Rummel, H. Hofmann, Phys.~Rev. {\bf E 64},  066126  (2001).

\bibitem{agb.ana:plb:98}
B.~K. Agrawal and A. Ansari, Phys.~Lett. {\bf B 421},  13  (1998).

\bibitem{hoh:pr:97}
H. Hofmann, Phys. Rep. {\bf 284 (4\&5)},  137  (1997).

\bibitem{grabert}
H. Grabert and U. Weiss, Phys.~Rev.~Lett. {\bf 53},  1787  (1984);
H. Grabert, P. Olschowski, and U. Weiss, Phys.~Rev. {\bf B 36},  1931  (1987).

\bibitem{hap.tap.bom:rmp:90}
P. H\"anggi, P. Talkner, and M. Borkovec, Rev. Mod. Phys. {\bf 62},  251
  (1990).

\bibitem{weissu}
U. Weiss, {\em Quantum Dissipative Systems} (World Scientific, Singapore,
  1993).

\bibitem{yos.kic.nak.noh:prc:00}
S.~K. You, C.~K. Kim, K. Nahm, and H.~S. Noh, Phys.~Rev. {\bf C 62},  045503
  (2000).

\bibitem{kleinerth}
H. Kleinert, {\em Pathintegrals in Quantum Mechanics, Statistics and Polymer
  Physics} (World Scientific, Singapore, 1990).

\bibitem{jaw.klh:cpl:87}
W.~Jahnke and H.~Kleinert, Chem.~Phys.~Lett. {\bf 137}, 162 (1987).

\bibitem{ruc.hoh:var-rate}
C. Rummel and H. Hofmann, nucl-th/0407092

\end{thebibliography}

\end{document}